\documentclass[prl,twocolumn,superscriptaddress]{revtex4}

\usepackage [dvips]{graphicx}                                   
\usepackage{amsmath}  
\usepackage{amssymb}    
\usepackage{mathrsfs}
\usepackage{color}

\def\olr{{\overline{\rho}}}

 \let\d\delta \let\th\theta \let\r\rho
\let\Si\Sigma

\newcommand{\beq}{\begin{equation}} 
\newcommand{\eeq}{\end{equation}}

\newcommand{\C}{\mathscr{C}}  
\newcommand{\ee}{\text{e}}     
\newcommand{\bx}{\text{\bf x}}
\newcommand{\F}{\mathscr{F}}
\newcommand{\by}{\text{\bf y}}      
\newcommand{\bk}{\text{\bf k}}
\newcommand{\p}{\partial}
\newcommand{\bnabla}{\boldsymbol{\nabla}}
\newcommand{\bj}{{\mathbf j}}
\newcommand{\bxi}{\boldsymbol{\xi}}
\newcommand{\bq}{\text{\bf q}}
\newcommand{\dd}{\text{d}}        
\newcommand{\bz}{\text{\bf z}}

\begin{document}

\title{Brownian dynamics: from glassy to trivial}
\author{Hugo Jacquin}
\affiliation{Laboratoire Mati\`ere et Syst\`emes Complexes, UMR 7057 CNRS/P7,
Universit\'e Paris Diderot -- Paris 7, 10 rue Alice Domon et L\'eonie 
Duquet, 75205 Paris cedex 13, France}
\author{Bongsoo Kim}
\affiliation{Department of Physics and Institute for Soft and Bio Matter Science, Changwon National University, Changwon 641-773, Korea}
\author{Kyozi Kawasaki}
\affiliation{Department of Physics, Faculty of Science, Kyushu University, Fukuoka 812-8581, Japan}
\author{Fr\'ed\'eric van Wijland}
\affiliation{Laboratoire Mati\`ere et Syst\`emes Complexes, UMR 7057 CNRS/P7,
Universit\'e Paris Diderot -- Paris 7, 10 rue Alice Domon et L\'eonie 
Duquet, 75205 Paris cedex 13, France}

\begin{abstract}
We endow a system of interacting particles with two distinct, local, Markovian 
and reversible microscopic dynamics. Using common field-theoretic techniques
used to investigate the presence of a glass transition,
we find that while the first, standard, 
dynamical rules lead to glassy behavior, the other one leads to a simple exponential 
relaxation towards 
equilibrium. This finding questions the intrinsic link that exists between the 
underlying, thermodynamical, energy 
landscape, and the dynamical rules with which this landscape is explored by 
the system. Our peculiar choice of dynamical rules offers the possibility of
a direct connection with replica theory, and our findings therefore call for a clarification
of the interplay between replica theory and the underlying dynamics of the system.
\end{abstract}

\maketitle

The microscopic phenomena driving the dynamical arrest in supercooled liquids 
and in glasses are still controversial. One line of thought, which originates 
in the work of Adam and Gibbs, pictures a glass-forming liquid as a system 
whose energy landscape complexity accounts for the slowing down of its 
dynamics. The original Adam-Gibbs~\cite{adamgibbs} theory relates viscosity --the momentum 
transport coefficient-- to configurational entropy. The more recent 
scenario due to
Kirkpatrick, Thirumalai, and Wolynes~\cite{RFOT_old}  is based on the 
study of the metastable configurations of the system and on the concept of 
nucleating entropic droplets. This is the 
Random-First-Order Theory (RFOT) scenario. 
The idea is that metastability arises from the many valleys of the energy landscape the system can be trapped in. The RFOT  scenario
gives a large set of quantitative predictions including non-mean field 
particle models \cite{real_RT}. These aspects of the physics of glasses were reviewed by 
Debenedetti and Stillinger~\cite{debenedettistillinger} and more recently by 
Biroli and Bouchaud~\cite{birolibouchaud}.

In another line of thought, no complex energy landscape needs to be invoked, 
and dynamically induced metastability alone is held responsible for the 
dynamics slowing down. This is a phenomenological approach in which at a 
coarse-grained scale local patches of activity are the only ingredients of 
the dynamics. This has led to the development of kinetically-constrained 
models (KCM) (see \cite{KCM} for a review), 
which have the advantage of lending 
themselves with greater ease than molecular models to both numerical 
experiments and analytical treatment. The hallmark of the corresponding
lattice models is the absence of any structure in the energy landscape (the 
equilibrium distribution is that of independent degrees of freedom). These 
somewhat simplistic descriptions are not directly built from the microscopics, 
and it is often argued that their glassy-like properties, like the existence 
of dynamical heterogeneities, are almost tautological, but recent 
works~\cite{speckchandler} have tried to bridge the gap from the 
microscopics to dynamic facilitation. 

A central concept in both approaches is that of metastability. A metastable
state can be viewed as an eigenstate of the evolution operator with a nonzero
but small relaxation rate. The existence of such states can be induced by the
dynamical evolution rules alone, as in KCM's, but physical intuition dictates
that, in realistic systems, these originate from the deepest structures of the
energy landscape. The method for capturing and counting metastable structures
from the statics alone was coined by Monasson \cite{Mo95}, and
further developed by Parisi and co-workers 
\cite{real_RT}, thus providing
the first first-principle calculation of some of the properties of
glasses. Their idea is to help polarize two attractively coupled copies of the system into one of these deep valleys and to investigate whether thermal fluctuations are enough to wash out any overlap between them. By contrast, capturing dynamically induced metastable 
structures is done by examining the overlap of the system properties in the 
course of a long time interval. Implementing this program was done by 
exploiting Ruelle's thermodynamic formalism first on KCM~\cite{us} and then in 
realistic systems~\cite{hedges,pitardlecomtevanwijland}.

Our purpose in this work is to show, as suggested in the past~\cite{kawasaki2003,miyazakireichman}, 
that a minute modification of the
evolution rules of a glass-forming liquid can lead to dramatic differences in 
the dynamics without affecting any of the static properties. A central 
difference with existing works is that ours will focus on interacting degrees of 
freedom, while KCM's, the system of oscillators studied earlier by Kawasaki 
and Kim~\cite{kawasakikim}, or the single oscillator of Whitelam and 
Garrahan~\cite{whitelamgarrahan}, are noninteracting degrees of freedom.

We consider a system of particles interacting via a two body potential $v$. For definiteness, and for the moment, we use a discretized version of our system in which configurations are labeled by a set of local occupation numbers $\C=\{n_j\}$ on a lattice. The energy of a configuration $\C$ is $E_P=\frac{1}{2}\sum_{i,j} n_i v(i-j) n_j$ and the equilibrium distribution is $P_\text{eq}(\C)=Z^{-1}\prod_{j}\frac{n_0^{n_j}}{n_j!}\ee^{-\beta E_P(\C)}$, 
which we shall also write in the form $P_\text{eq}(\C)=Z^{-1}\ee^{-\beta E(\C)}$ 
where $E(\C)$ comprises an energy component and an entropy component, and where $n_0$ is the average occupation number and $\beta$ is the inverse temperature. It is known since the work of De Dominicis {\it et al.}~\cite{dedominicisorlandlainee} that picking transition rates between configurations $\C$ and $\C'$ such that $W(\C\to\C')=P_\text{eq}(\C')$ leads to a master equation whose spectrum is made of the zero eigenvalue of the equilibrium state, and a unit eigenvalue (with of course a very large degeneracy). Under such conditions, the dynamics, and thus all time correlations, will 
exponentially relax to equilibrium at unit rate. A more physical choice proposed by Koper and Hilhorst~\cite{koperhilhorst} is to weight transition between states according to an Arrhenius-like factor, namely $W(\C\to\C')=\ee^{-\frac{\beta}{2}(E(\C)-E(\C'))}$. Then again, the techniques of Koper and Hilhorst allow one to prove that no quasi-degenerate eigenvalues with the equilibrium state will exist, and fast exponential relaxation will occur to the equilibrium state. In the absence of disorder, no further average of the relaxation rate over some frozen-in degrees of freedom is needed and thus no slowing down (in the form of stretched exponentials or power law behavior) will emerge. These examples alone could suffice to illustrate that albeit the energy landscape may feature a wealth of wells and other (mechanically) metastable structures, dynamics can overcome these and show no sign of glassiness whatsoever. Note however that in these somewhat academic examples, the dynamics --not the statics 
though-- 
possess mean-field features 
since all states are dynamically connected,
 which involves non local displacements of particles. Thus we would like to bring forth an example of dynamics in which the latter mean-field feature has been replaced with local dynamics. To illustrate our point, we will compare two possible dynamics for our system of interacting particles, which are both local in space, time-reversible and Markovian. We begin by restricting possible transitions between  states in which a single particle has hopped from one site to one of its nearest neighbors. We define the hopping rate of a particle at site $i$ to site $j$ by 
\begin{align}\label{rates}
\left\{ \begin{array}{l l}
n_i e^{- \beta  \Delta E_p / 2} & \text{for Langevin dynamics} \\
n_0 \sqrt{\frac{n_i}{n_{j}+1}} e^{- \beta \Delta E_p/2} & \text{for Model $B$
  dynamics} \\
\end{array} \right.
\end{align}
where $E_p=\frac 12 \sum_{<i,j>} n_i v(i-j) n_j$ ($<i,j> $ denoting
the pairs of nearest neighbor sites) and $\Delta E_p$ is the 
potential energy variation due to the particle hopping. The {\it Langevin dynamics} terminology is due to the observation that in the continuum coarse-grained limit, as discussed in \cite{lefevrebiroli}, these evolution rules describe a set of Brownian particles interacting via the two-body potential $v$. We shall soon provide the reader with an intuitive explanation of the {\it Model B} terminology for the rates in the second line of \eqref{rates}. Starting from these lattice dynamics, 
standard manipulations \cite{doi_peliti} allow one to write down a path integral 
representation for the two dynamics. In the coarse-grained limit, they lead to path integral representations for the local density field $\rho(\bx,t)$, which, in turn, can be phrased in terms of a Langevin equation for this local density. Instead of phrasing out these technical steps, we adopt an equivalent but more phenomenological approach. For a system with average density $\rho_0$, the equilibrium distribution for the density modes $\rho(\bx)$ is $P_{\text{eq}}[\rho] \propto \ee^{- \beta \F[\rho]}$, where the density 
dependent effective free energy has the well-known expression
\begin{align}
&
\beta\F[\rho] = \int_\bx \left[  \rho(\bx) \ln  \frac{\rho(\bx)}{\rho_0} 
 - \rho(\bx) \right] \nonumber \\
& 
\phantom{\beta \F[\rho] =} + \frac 12 \int_{\bx,\by} \rho(\bx) \beta 
v(\bx-\by) \rho(\by),
\end{align}
in which the first term in the rhs is of entropic origin, and the second one 
is the potential interaction energy. The statics alone tell us about the equilibrium structure factor $S_\bk$ that we shall later need. It should be noted that beyond the replica approach, attempts have been made at connecting the minima of $\F$ to the supercooled state~\cite{kaurdas}.

From a dynamical point of view, the particle density must be locally conserved, which constrains it to evolve according to a continuity equation 
\begin{equation}
\p_t\rho+\bnabla\cdot \bj=0.
\end{equation}
Our two choices of dynamics lie in the specific density dependence of the 
particle current $\bj$. 
We begin with a standard choice in which the particle 
current is given by
\begin{equation}\begin{split}\label{current_dean}
\bj_L& =-\rho\bnabla\frac{\delta \F[\rho]}{\delta \rho}+\sqrt{2T\rho} \, \bxi \\
& = - T \bnabla \rho - \rho \int \bnabla v \rho + \sqrt{2T\rho} \, \bxi \ .
\end{split}\end{equation}
where the parameter $\beta \equiv T^{-1}$ is the inverse temperature of 
the thermostat, and $\bxi$ is a vector whose components are independent white noises with 
unit correlations 
$\langle\xi^\alpha(\bx,t)\xi^\beta(\bx',t')\rangle=\delta^{\alpha\beta}\delta^{(d)}
(\bx-\bx')\delta(t-t')$, where $d$ is the space dimension. 
In an explicit form, the deterministic part of 
$\bj_L$ features a Fickian diffusion term $-T\bnabla\rho$ and a driving term 
$\rho { \mathbf F}$, where ${\mathbf F}(\bx,t)=-\int_\by \bnabla v(\bx-\by)
\rho(\by,t)$ is the local fluctuating force field acting on particles located 
around $\bx$. We index the particle current with the letter $L$ because this 
very expression of $\bj_L$ was shown by Dean~\cite{dean} to exactly encode the 
dynamics of a fluid of interacting particles individually evolving through an overdamped 
Langevin equation.  Once the deterministic part of $\bj_L$ is set, 
time-reversibility (or detailed balance) forces the specific $\sqrt{2T\rho}$ 
density dependence of the noise strength. The expression of ${\bf j}_L$ exactly encodes the rates in the first line of \eqref{rates}. 
Our second choice of dynamics for the collective modes 
is defined by another expression for the particle current, namely
\begin{equation} \begin{split}
\bj_B & = - \rho_0 \bnabla\frac{\delta \F[\rho]}{\delta \rho} + \sqrt{2 T\rho_0} \, \bxi  \ ,\\
& = - T \rho_0 \frac{\bnabla \rho}{\rho} - \rho_0 \int \bnabla v \rho + \sqrt{2T\rho_0} \, \bxi
\end{split}
\label{current_modelB}
\end{equation}
where the index $B$ is reminiscent of model $B$ dynamics of the Hohenberg and
Halperin classification~\cite{hohenberg}, as already defined for discussion
purposes in \cite{dean}. The difference with the standard model B
dynamics is of course that the underlying Hamiltonian is not that of a standard $\phi^4$ theory, a situation that has been studied in a related context in \cite{geissler}.
Note that in \eqref{current_modelB} the coupling to
the thermal bath is now independent of the local density (noise is additive instead
of being multiplicative as in \eqref{current_dean}). Dynamics $B$ can be
obtained from dynamics $L$ by thinking the fluctuations of the coupling to the
thermostat have been turned off, meaning a stronger coupling to the thermostat controlling particle source. 
We emphasize that the detailed balance
property is satisfied also in these alternative dynamics, which ensures that in
principle both $\bj_L$ and $\bj_B$ drive the system towards the same Gibbs
equilibrium state $P_\text{eq}[\rho]$. Model B dynamics can be shown to derive from the rates in the second line of \eqref{rates}.\\

We already see that the above two different forms of currents points to fundamental 
 dynamic distinctions in
the two types ($L$ and $B$) of dynamic evolutions towards the same equilibrium state.
Quantitative consequences of these distinction are detailed below.
While the ideal gas free energy leads to the simple Fickian diffusion in $L$ dynamics
with constant diffusion constant, in $B$ dynamics it leads to the non-Fickian diffusion 
with the diffusion constant inversely dependent on the local density. 
While in $L$ dynamics the Gaussian particle interaction leads to the density nonlinearity which can drive the system glassy (forming cages) with increasing density, 
the same Gaussian interaction merely gives the linear dependence on the density.
The entire (nonpolynomial) nonlinearity comes from the ideal-gas part of the free energy in 
the case of $B$ dynamics, which is somewhat unexpected to give rise to a glassy behavior.

Our goal here is not to come up with a new approximation for the Langevin 
dynamics. A huge body of numerical and analytical literature is devoted to 
extracting signatures of glassiness from these dynamical rules~\cite{das}. We 
take them for granted. Instead, our purpose is to conduct an analytical study 
of our model $B$ dynamics in which we will argue that no sign of dynamical 
arrest can be found. We examine the time evolution equation for the 
density-density correlation function (otherwise called the intermediate 
scattering function). If a glassy behavior is to be found,
then we must obtain a non vanishing long-time limit of this correlation 
function, indicating ergodicity breaking, or, if not, we should at least 
expect signs for a slow relaxation. We shall now argue that 
using standard field-theoretic techniques, model $B$ dynamics 
shows no sign of anomalous density-density correlations.
Without entering calculational details, we now sketch how to obtain an evolution equation for the density-density correlation function 
$G_{\rho\rho}(\bk,t)$. Keeping calculational details to a low level, we begin by writing the Langevin equation in its path-integral representation 
{\it \`a la } Martin-Siggia-Rose-Janssen-De Dominicis in which an additional response field (conjugate to the noise) is introduced:
\begin{equation}
S[\bar{\rho},\rho]=\int_{\bx,t}\left[\bar{\rho}\left(\p_t\rho-\rho_0\bnabla^2\frac{\delta \F}{\delta\rho}\right)-T\rho_0(\bnabla\bar{\rho})^2\right]
\end{equation}
The non-quadratic part of the action is contained in $-\int_{\bx,t} T\rho_0\big(\bnabla^2\bar{\rho}\big)\,\left[\ln\left(1+\frac{\delta\rho}{\rho_0}\right)-\frac{\delta\rho}{\rho_0}\right]$, where $\delta\rho$ denotes the density fluctuation field. If one denotes by $\Sigma_{\bar{\rho}\bar{\rho}}$ the loop corrections to the $\bar{\rho}-\bar{\rho}$ vertex function, then one can prove the following evolution equation for $G_{\rho\rho}$:
\begin{equation}\label{mctlike}
\left(\p_t+\frac{Tk^2}{S(k)}\right)G_{\rho\rho}(\bk,t)=\int_0^t\dd s ~ \frac{\Sigma_{\bar{\rho}\bar{\rho}}(\bk,t-s)}{T\rho_0 k^2}\p_s G_{\rho\rho}(\bk,s)
\end{equation}
This is the first important result of our work. In order to derive it, one begins by writing the Schwinger-Dyson equation for the two-point correlation functions, in the form $G_0^{-1}G=\mathbf{1}-\Sigma G$, where $G$ is the $2\times 2$ matrix of two-point correlations for the $\bar{\rho},\rho$ fields, $\Sigma$ is the related matrix of vertex functions, and $G_0$ is the Gaussian part of $G$.  Then we use the fluctuation-dissipation theorem, which  is manifested in the following reversibility symmetry of the action
\begin{align}\label{TRJ}
\left\{ \begin{array}{l l}
\delta \rho(t) & \to \delta \rho(-t) \\
\displaystyle \bnabla^2\bar{\rho}(t) & \displaystyle \to \bnabla^2\bar{\rho}(-t) +  \frac{\beta}{\rho_0} \p_t\rho
\end{array} \right.
\end{align}
Not only correlations between the fields are linked to each other, but also two-point vertex functions are related when time is reversed~: $\Sigma_{\bar{\rho}\rho}(t,t')=\Sigma_{\bar{\rho}\rho}(-t,-t')+\frac{1}{\rho_0 Tk^2}\frac{\dd}{\dd t}\Sigma_{\bar{\rho}\bar{\rho}}(t,t')$, where the space Fourier variables were omitted. Once this relation is inserted into the Schwinger-Dyson equation, using an integration by parts and the fact that $\Sigma_{\bar{\rho}\rho}$ is causal, one finally arrives at \eqref{mctlike}. The form of that equation is well-known to the practioners of the mode-coupling theory, but here, and so far, no approximations were made. No such equation could be written within the framework of the Langevin dynamics without further approximation, because, due to the multiplicative nature of the noise, the physical response function is not simply connected to the two point $\bar{\rho}-\rho$ correlations which must inevitably and independently come into play. The 
vertex function $\Sigma_{\bar{\rho}\bar{\rho}}$ which plays the role 
of a memory kernel is now the subject of our concern. It should be remarked that it can be written as a functional of the two correlation functions $G_{\rho\rho}$ and 
$G_{\rho\bar{\rho}}$. The latter, by virtue of the fluctuation-dissipation theorem, is proportional to the time derivative of $G_{\rho\rho}$, hence $\Sigma_{\bar{\rho}\bar{\rho}}$ is a functional of $G_{\rho\rho}$ only. The first intriguing remark is that the interaction potential between particles does not explicitly enter the interaction vertices of the field theory --these are due to the ideal gas contribution only-- and in fact it only appears through the static structure factor $S_\bk$ that governs the linear relaxation (the latter being a well-behaved functions mildly deviating from unity for many systems 
that are known to exhibit glassy features (such as  harmonic spheres). 
This means that the functional $\Sigma_{\bar{\rho}\bar{\rho}}[G_{\rho\rho}]$ is the same with or without interactions between particles. Second, up to an overall proportionality factor, the expression of $\Sigma_{\bar{\rho}\bar{\rho}}[G_{\rho\rho}]$ would be the same for a more artificial model A dynamics (without local particle conservation) as it is for our model B dynamics. 
This, again, is a consequence of the fluctuation-dissipation theorem.

\begin{figure}[h]
\includegraphics[width=8cm]{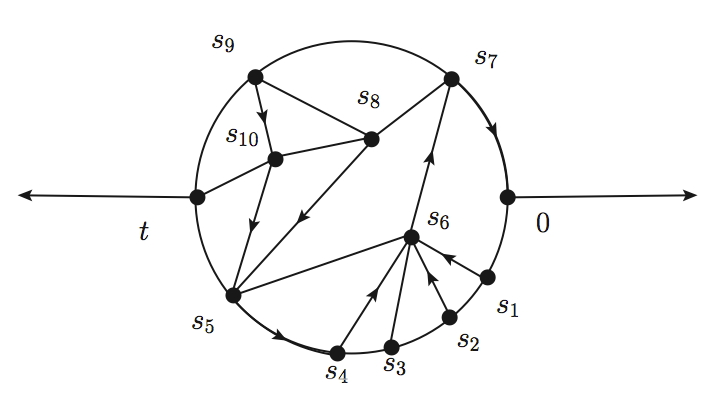}
\caption{An arbitrary diagram contributing to $\Sigma_{\bar{\rho}\bar{\rho}}(t)$ involving ten internal vertices. Time increases along an arrowed-leg. A simple leg carries a correlation function and a causal arrowed one carries its time derivative due to the fluctuation-dissipation relation.}
\label{diag-arbit}                                                                                                                                                             
\end{figure}

We now explain why an Ergodic -- Non-Ergodic (ENE) transition must be ruled out. We assume that at large times $G_{\rho\rho}(\bk,t\to\infty)=\rho_0 S_\bk f_\bk$ where the nonergodicity parameter is {\it a priori} nonzero. Before we consider an arbitrary 2PI diagram, a subclass of diagrams made of the watermelons is of specific interest. The simplest one, already examined in \cite{FDT} in a different context, diverges for any nonzero $f_\bq$. A watermelon with three internal lines may well converge (as in \cite{geissler}), the previous diagram will nevertheless always win over. Consider now an arbitrary 2PI diagram contributing to $\Sigma_{\bar{\rho}\bar{\rho}}(t)$, such as the one shown in figure \ref{diag-arbit}. The most divergent part of the diagram for $t\to\infty$ comes from setting the non-arrowed legs to $\rho_0 S_\bq f_\bq$. For a diagram with $n$ internal vertices, there are $n$ arrowed legs which carry the time-derivative of $G_{\rho\rho}$. One then performs time integrations from the most-nested vertex outwards (in figure \ref{diag-arbit} this would start with $s_1$, then $s_3, s_2, s_9, s_{10}, s_8, s_5, s_4, s_6$ and finaly $s_7$). What remains for the corresponding diagram is a product of factors $\rho_0 S_\bq (1-f_\bq)$ over the momenta $\bq$ carried by each arrowed legs, times a product of $\rho_0 S_\bq f_\bq$ over the momenta carried by the non-
arrowed legs. The general trend for an $n$-vertex diagram is $(1-f)^n f^{2+m}$, where $m\geq 1$. 
The resulting equation, because of the leading order 2 in $G_{\rho\rho}$ contribution, will not have a nonzero solution. This seems to rule out the scenario of an ENE transition, but one cannot exclude a scenario in which an infinite resummation of all diagrams (or even some subclass of these) could restore an ENE transition. This kind of scenario was described in a closely related context by Jacquin and Zamponi~\cite{jacquinzamponi}.

However one would then like to know how $G_{\rho\rho}$ relaxes to 0 with time. This behavior will be contained in the memory kernel $\Sigma_{\bar{\rho}\bar{\rho}}$ that we now study. Following the lines of \cite{KKJW14}, we introduce an auxiliary field $\th$, that contains all the non-linearities in the density field:
\beq
\th(x,t) =f(\delta \rho(x,t))
\equiv \ln \left( 1 + \frac{\d \rho(x,t)}{\r_0} \right) - \frac{\d \rho(x,t)}{\r_0} \ .
\eeq
This field plays a crucial role in analyzing the Dean-Kawasaki dynamics in order to linearize the time-reversal symmetry, but we introduce it here
merely to perform a resummation of a large number of diagrams at once.
Following the first steps in \cite{KKJW14} by projecting the equation of motion for the $\delta\rho$ field onto $ \delta\rho$ and $\theta$, 
leads, respectively to
\begin{equation}\label{addit}\begin{split}
\p_t G_{\rho\rho}+Tk^2(1+\beta\rho_0 v(k))G_{\rho\rho}=-\rho_0 T k^2 G_{\rho\theta}\\
\p_t G_{\rho\theta}+{Tk^2(1+\beta\rho_0 v(k))}G_{\rho\theta}=-\rho_0 T k^2 G_{\theta\theta}
\end{split}\end{equation}
supplemented with the equal time equilibrium values
$G_{\rho\rho}(\bk,0)=\rho_0 S_\bk$, 
$G_{\rho\theta}(\bk,0)=1-S_\bk(1+\beta\rho_0 v(k))$ and $G_{\theta\theta}(\bk,0)=-\frac{1}{\rho_0}(1+\beta\rho_0 v(k))G_{\rho\theta}(\bk,0)$. Equations \eqref{addit} and \eqref{mctlike} give three equations for the four unknowns, $G_{\r\r}$, $G_{\r\th}$, $G_{\th\th}$ and $\Si_{\olr\olr}$. As in the previous attempts, an expression for $\Sigma_{\bar{\rho}\bar{\rho}}$ in terms of the various $G$ 
correlations must be found to arrive at a closed set of equations. The simplest closure scheme that keeps all the terms in the expansion of $\theta=f(\delta\rho)$  is to carry out a cumulant expansion $\langle\ee^{\int\rho_0 Tk^2\bar{\rho}\theta}\rangle_0\simeq\ee^{\frac 12 \langle\left(\int\rho_0 Tk^2\bar{\rho}\theta\right)^2\rangle_0}$, where the $0$ index denotes an average with respect to the quadratic action (at fixed $G$'s). This immediately leads to 
\begin{equation}\label{approx2}
\Sigma_{\bar{\rho}\bar{\rho}}=-(T\rho_0 k^2)^2 G_{\theta\theta}
\end{equation}
where the $G_{\theta\theta}$ function that appears in \eqref{approx2} is to be understood as a functional of $G_{\rho\rho}$ as given by Wick's theorem. 

 It is a simple exercise to write \eqref{addit} and \eqref{mctlike}, combined with \eqref{approx2} in Laplace form, and to solve for $G_{\r\r}$.
We find that finite relaxation rates are found by this procedure, and no particular slowing down is identified. For this particular truncation scheme the two rates for the exponential decay are $Tk^2(1+\beta\rho_0 v(k))$ and $Tk^2(S_\bk+1+S_\bk\beta\rho_0 v(k))^2/4S_\bk$ and they carry qualitatively the same physical meaning. It is also important to note that the ideal gas limit $S_\bk\to 1$ and $v(\bk)\to 0$ poses no particular problem. We therefore conclude that away from any thermodynamic phase transition, the structure factor displays no specific singularity and thus the relaxation rate remains finite again, within this approximation.

Improvements would consist in going to higher cumulants, which would involve the cross correlation $G_{\theta\bar{\rho}}$ which can itself, by the FDT, be expressed in terms of $G_{\theta\theta}$ and $G_{\theta\rho}$ ($G_{\theta\bar{\rho}}(t)=\Theta(t)(K G_{\rho\theta}+G_{\theta\theta}))$). We believe that our conclusions would not be changed.

The self-consistent equation for the nonergodicity parameter $f_\bk$  is independent of the details of the dynamics, for as long as the dynamics has additive noise (and drives the system to its equilibrium distribution). This means that if an ENE transition existed, the equation for $f_\bk$ would be identical as that deduced from a model A type of dynamics. Though the reasoning was performed on a different system, Crisanti has shown in \cite{crisanti} that a model A self-
consistent equation for $f_\bk$ (as obtained from the dynamic MCT approximation) is identical to the one a static one-step replica symmetry breaking calculation (as in \cite{jacquinzamponi}) would yield. 
In passing, it is interesting to note that, by extrapolation, Crisanti's argument would connect replica calculations to our model B dynamics rather than to the physical Langevin Dean-Kawasaki dynamics.  
It is very puzzling to observe that Crisanti's argument, combined with
the replicated hyper-netted chain approximation of replica theory \cite{MP96} seems to prove that an ENE phase transition could be detected via the model B dynamics, 
while all our computations tend to show the contrary. 
In order to reconcile these apparently conflicting
observations, it would be a step forward to understand
whether there is any flaw either in Crisanti's argument when applied to the case of structural glasses, or in the resummation needed to be performed on the dynamics side.
Interestingly, the same kind of controversy was initiated by the work of \cite{geissler}, where MCT-like approximations and replica theory seemed
to predict the existence of a transition, while numerical simulations seemed to prove the contrary, although to our knowledge the argument has not been fully 
settled \cite{dispute}. 
In this particular case, a very crude approximation for the dynamics performed better when compared to the simulation
then a Mode-Coupling one, which predicted a spurious transition.

In retrospect, the the dynamical rules considered so far are very similar to 
each-other. Only local moves are involved, they both preserve particle 
conservation and for both time-reversibility is encoded in the dynamics. 
Closer inspection, however, shows potentially relevant differences. Inspired by earlier works~\cite{pitardlecomtevanwijland}, we would like to examine the the rate at which the system leaves a configuration characterized by the density profile $\rho$. This escape rate, as it is termed in the theory of Markov processes, can be obtained by summing the transition rates in \eqref{rates} over possible target configurations after the continuum limit has been taken. Of course simple particle hops contribute to this escape rate. Here we wish to focus only on the interaction-potential dependent part of the escape rate and on its dependence on the instantaneous local density profile, which we denote by $r[\rho]$.
For Langevin dynamics, this escape rate reads 
\begin{align}\label{rateL}
r_L[\rho] = & ~ \frac{\beta^2}{4}\underset{\bx,\by,\bz}{\int}\nabla v(\bx-\by)
\cdot \nabla v(\bx-\bz) \rho(\bx,t)\rho(\by,t)\rho(\bz,t)\nonumber\\
& -\frac{\beta}{2}\underset{{\bx,\by}}{\int}\rho(\bx,t)\nabla^2 v(\bx-\by)
\rho(\by,t)
\end{align}
The average rate, as can be seen in the first term in the right hand side 
of \eqref{rateL}, involves two and, more importantly three-body correlations. 
The knowledge of these correlations is necessary for the system to decide to 
which location of phase space it is going to evolve. Turning now to the 
model $B$ dynamics, we find that
\begin{align}
r_B[\rho] = & ~ \frac{\beta^2\rho_0}{4}\underset{\bx,\by,\bz}{\int}\bnabla 
v(\bx-\by) \rho(\by,t)\cdot \bnabla v(\bx-\bz) \rho(\bz,t)\nonumber\\
& +\beta\rho_0
\underset{\bx,\by}{\int}\frac{\bnabla\rho}{\rho}(\bx,t) \cdot \bnabla v(\bx-\by)
\rho(\by,t)
\end{align}
so that on average the system requires no more than two-point correlations to 
proceed with its evolution. The Langevin dynamics requires a more refined 
knowledge of the local organization than model $B$ dynamics. We speculate 
that we may relate this mathematical observation to the physical picture of 
local caging. It is a well-known fact that the pair correlation function of 
a glass is no different from that of a liquid, but recent 
works~\cite{mayermiyazakireichman, coslovich} suggest that triplet 
correlations (and higher order correlations) may behave differently. 

Our result stands out as one in which the dynamics of a system of interacting particles is sufficiently simple to be analyzed in terms of two-point functions. 
Yet it is based on a coarse-grained formulation in which the collective
density modes are treated as smoothly varying function of space. For softly 
repulsive potentials, the system will fall into a jammed
state as the density is increased. Our approach is thus bound to reach its
limitations at high densities. However, in the usual glassy regime where 
ergodicity is maintained, a standard two-step relaxation will never be
observed with our choice of dynamics.

A physical picture supporting our results was given by Cates~\cite{cates}. It relies on identifying a one-particle Langevin dynamics whose description in terms of collective modes is that of our model B. Such dynamics will differ from the standard Langevin dynamics in that the particle's mobility, instead of being constant, will be proportional to the reciprocal of the local density. With this picture in mind, one can ask Kramers' question, in a one-dimensional setting, and find the typical time it takes to get across a potential barrier. In the standard derivation this time involves the ratio of the local equilibrium densities at the bottom of the potential well and up at the top of the barrier. However in the model B picture the local mobility exactly compensates for the rarity/propensity of hopping and the local dynamics thus become largely insensitive to the details of the potential energy landscape.

It would in fact be an interesting step to describe the individual effective particle dynamics behind our model B  dynamics, say in terms of an effective Langevin equation for the particles' positions. It could then provide an efficient method for simulating systems plagued with slow dynamics, allowing the system to bypass mechanically metastable states and force them to reach equilibrium.

From a broader perspective, our calculation points 
to the necessity of establishing criteria permitting to relate static 
energy-landscape-based considerations, to kinetic properties. A naive way 
to ask the question would be: What are the generic conditions on the 
effective diffusion constant coupling the energy gradient to the thermal 
bath that allow for a reliable correspondence between minima of the energy 
landscape, and metastable states? This question echoes a recent work 
\cite{zampo_krzakala} in which it is shown
that one could map a subclass of KCM models onto a subclass of spin-glasses, uncovering
a non-trivial behavior of appropriately defined static quantities in these KCMs, which could mirror non-trivial dynamical transitions in the corresponding spin-glass systems. Once
again, the traditional separation between static and dynamic framework does
thus not seem to hold anymore, calling for a more refined treatment of the
interplay between the two.

A quantitative characterization of the dynamic complexity, as expressed for
example by the Lyapunov spectrum of our two 
dynamics could uncover the fundamental difference that exist between the two. 
More generally, what is the signature of an ergodicity breaking transition on 
the Lyapunov spectrum? These are ongoing investigations.

We warmly acknowledge discussions with Matthias Fuchs, Vivien
Lecomte, Kunimasa Miyazaki, Estelle Pitard, Peter Sollich, Grzegorz Szamel and
Francesco Zamponi. H. J. acknowledges funding from a fondation CFM-JP Aguilar grant and from
ERC grant OUTEFLUCOP.
BK gratefully acknowledges a support by the research unit FOR 1394 'Nonlinear response to probe vitrification' funded by the DFG and the hospitality during his sabbatical  
at the University of Konstanz.
BK and KK are also supported by the Basic Science Research Program through the NRF funded by the Ministry of Education, Science, and Technology (Grant No. 2011-0009510).


\bibliography{scibib}
\bibliographystyle{unsrt}

\end{document}